\documentclass[runningheads]{llncs}

\usepackage[T1]{fontenc}

\usepackage{enumitem} 
\usepackage{graphicx}
\usepackage{url}
\usepackage[caption=false]{subfig}
\usepackage{hyperref}
\usepackage{makecell}
\usepackage{etoolbox}
\usepackage{xcolor}
\usepackage{multirow}
\usepackage[misc]{ifsym}

\usepackage{cite}
\usepackage{xcolor}
\definecolor{boxgrey}{HTML}{F3F3F3}

\newcommand{\hlbox}[2]{
  \begin{center}
    \fcolorbox{white}{boxgrey}{
      \parbox{0.9\columnwidth}{\noindent \textbf{#1}. \textit{#2}}
    }
  \end{center}
}
\begin{document}

\title{Knowledge is Power, Understanding is Impact: Utility and Beyond Goals, Explanation Quality, and Fairness in Path Reasoning Recommendation}
\titlerunning{Utility and Beyond, Explanations, Fairness in Path Reasoning}

\author{Giacomo Balloccu\inst{1}\orcidID{0000-0002-6857-7709} \and
Ludovico Boratto\inst{1}\orcidID{0000-0002-6053-3015} \and
Christian Cancedda\inst{2}\orcidID{0000-0002-8206-3181} \and
Gianni Fenu\inst{1}\orcidID{0000-0003-4668-2476} \and
Mirko Marras\inst{1}\textsuperscript{(\Letter)}\orcidID{0000-0003-1989-6057}}

\authorrunning{G. Balloccu et al.}

\institute{University of Cagliari, Cagliari, Italy\\ 
\email{giacomo.balloccu@acm.org}, \email{ludovico.boratto@acm.org}\\
\email{fenu@unica.it}, \email{mirko.marras@acm.org} 
\and Polytechnic University of Turin, Turin, Italy\\
\email{christian.cancedda@studenti.polito.it}}

\maketitle            

\begin{abstract}
Path reasoning is a notable recommendation approach that models high-order user-product relations, based on a Knowledge Graph (KG). 
This approach can extract reasoning paths between recommended products and already experienced products and, then, turn such paths into textual explanations for the user. 
Unfortunately, evaluation protocols in this field appear heterogeneous and limited, making it hard to contextualize the impact of the existing methods.
In this paper, we replicated three state-of-the-art relevant path reasoning recommendation methods proposed in top-tier conferences. 
Under a common evaluation protocol, based on two public data sets and in comparison with other knowledge-aware methods, we then studied 
the extent to which they meet recommendation utility and beyond objectives, explanation quality, and consumer and provider fairness. 
Our study provides a picture of the progress in this field, highlighting open issues and future directions. 
Source code: \url{https://github.com/giacoballoccu/rep-path-reasoning-recsys}.
\keywords{Recommender Systems, Knowledge Graphs, Replicability.}
\end{abstract}

\section{Introduction}
Recommender systems (RS) are a popular strategy to enable personalized users' experience \cite{ricci}.
Historical data (e.g., browsing activity and ratings) and product characteristics (e.g., title and description) are well-recognized data sources to train RSs. 
Product information is often augmented with \emph{Knowledge Graphs} (KGs)~\cite{cao-etal-2018-neural,10.1145/2926718}. 
These KGs include \emph{entities} (e.g., users, movies, actors) and \emph{relations} between entities (e.g., an actor starred a movie). 
Integrating KGs within RSs has led to a gain in recommendation utility \cite{kgat, ripple}, especially under sparse data and cold-start scenarios \cite{9251221}. 
Their inclusion is essential to make RS explainable and turn recommendation into a more transparent social process \cite{explainable-recsys-survey,Tintarev2007}.

Notable recommendation methods based on KGs include \emph{path reasoning methods}
~\cite{cfkg,10.1609/aaai.v33i01.33015329,ma2019jointly,kprn,ni-etal-2019-justifying,musto2021generating,pgpr,Song2019EkarAE,leveraging-demostrations}. 
To guide RS training, they rely on paths that model high-order relations between users and products in the KG,
and identify those deemed as relevant between already experienced products and products to recommend.  
Such paths are also used to create explanations, through explanation templates or text generation. 
In the movie domain, the path “user$_1$ watched movie$_1$ directed director$_1$ directed$^{-1}$ movie$_2$” might lead to 
the template-based explanation “movie$_2$ is recommended to you because you watched movie$_1$ also directed by director$_1$”.
Path reasoning methods are in contrast to regularization methods, which weight product characteristics based on their importance for a given recommendation
but do not provide any explanation~\cite{cke,deepcon,sequentialmemorykg,he2020mining, ripple, kgat}. 

An abundance of KGs were proposed for recommendation, along with path reasoning methods, to produce both recommendations and explanations~\cite{arrieta2019explainable}.
However, evaluation protocols were heterogeneous (e.g., different train-test splits) and limited to a narrowed set of evaluation data sets and metrics. 
Prior works often showed that a novel method led to a higher recommendation utility, compared to (non) knowledge-aware baselines. 
None of the them deeply analyzed beyond utility goals (e.g., coverage, serendipity) nor monitored consumer (i.e., end users) and provider fairness.
Hence, it remains unclear whether path reasoning methods emphasize any trade-off between goals unexplored so far. 
Being the landscape convoluted and polarized to utility, there is a need for a common evaluation ground to understand how and when each method can be adopted.

In this paper, we conduct a replicability study (different team and experimental setup) on unexplored evaluation perspectives relevant to path reasoning methods. 
In a first step, we scanned the proceedings of top-tier conferences and journals, identifying seven relevant papers. 
We tried to replicate the original methods based on the released source code, but only three of them were replicable. 
In a second step, we defined a common evaluation protocol, including two public data sets (movies; music), two sensitive attributes (gender; age),
and sixteen metrics pertaining to four perspectives (recommendation utility; beyond utility goals; explanation quality; fairness). 
We evaluated path reasoning methods under this protocol and compared them against other knowledge-aware methods. 
Results reveal that, despite of an often similar utility, path reasoning methods differ in the way they meet other recommendation goals. 
Our study calls for a broader evaluation of these methods and a more responsible adoption.  

\section{Research Methodology}
In this section, we describe the collection process for path reasoning methods, the steps for their replication, and the common evaluation protocol.

\subsection{Papers Collection}
To collect existing path reasoning methods, we systematically scanned the recent proceedings of top-tier information retrieval events
(CIKM, ECIR, ECML-PKDD, FAccT, KDD, RecSys, SIGIR, WSDM, WWW, UMAP) 
and journals edited by top-tier publishers (ACM, Elsevier, IEEE, Springer). 
The adopted keywords combined a technical term between ``\emph{path reasoning recommender systems}'' and ``\emph{explainable recommender system}''
and a non-technical term between ``\emph{explainable AI}'' and ``\emph{knowledge enabled AI}''. 
We marked a paper as relevant if (a) it addressed recommendation, (b) it proposed a KG-based method, and (c) the method could produce reasoning paths. 
Papers on other domains or tasks, e.g., non-personalized rankings or mere entity prediction tasks (w/o any recommendation) were excluded. 
We also excluded knowledge-aware methods unable to yield reasoning paths, although we will use some representatives of this class for comparison.
Seven relevant papers were selected for our study (Table \ref{tab:papers}).

We attempted to replicate the method of each relevant paper, relying as much as possible on the original source code.
To obtain it, we first tried to search for the source code repository into the original paper and on the Web.
As a last resort, we sent an e-mail to the original authors. 
We considered a method to be replicable in case a fully working version of the source code was obtained 
and needed minor changes to accept another data set and extract recommendations (and reasoning paths). 
Three out of the seven relevant papers were replicable with a reasonable effort. 
As per the non-replicable ones, three did not provide any source code\footnote{
Note that the source code of these papers might appear soon online as an effect of our e-mails to the original authors. We leave their replication as a future work.  
}. The other one included unavailable external dependencies \cite{10.1162/dint_a_00013}. 

\begin{table}[!t]
\caption{Path reasoning methods deemed as relevant in our study.}
\label{tab:papers}
\setlength{\tabcolsep}{6pt}
\resizebox{1\linewidth}{!}{
\begin{tabular}{llc|ccccc}
\hline
\textbf{Method} & \textbf{Year} & \textbf{Status}$^1$ & \multicolumn{5}{c}{\textbf{Experimental Setting}} \\
& & & \textbf{Data Sets}$^2$ & \textbf{Split Size}$^3$ & \textbf{Split Method}$^4$ & \textbf{Recommendation$^5$} & \textbf{Explanation$^5$} \\
\hline
\texttt{PGPR} \cite{pgpr}                       & 2019 & $RE$               & $AZ$              & $70$-$00$-$30$ & $Rand$ &  NDCG, R, HR, P & -                         \\ 
\texttt{EKAR}  \cite{Song2019EkarAE}            & 2019 & $\overline{RE}$    & $ML$, $LFM$, $DB$  & $60$-$20$-$20$ & $Rand$ &  NDCG, HR                  & -                         \\ 
\texttt{CAFE} \cite{cafe}                       & 2020 & $RE$               & $AZ$              & $70$-$00$-$30$ & $Rand$ &  NDCG, R, HR, P & -                         \\ 
\texttt{UCPR} \cite{usercentric}                & 2021 & $RE$               & $ML$, $AZ$              & $60$-$20$-$20$ & $Rand$ &  NDCG, R, HR, P & PPC\\
\texttt{MLR} \cite{10.1145/3485447.3512083}     & 2022 & $R\overline{E}$    & $AZ$              & $70$-$00$-$30$ & $Rand$ &  NDCG, R, HR, P & -                         \\ 
\texttt{PLM-Rec} \cite{10.1145/3485447.3511937} & 2022 & $\overline{RE}$    & $AZ$              & $60$-$20$-$20$ & $Time$ &  NDCG, R, HR, P & -                         \\ 
\texttt{TAPR} \cite{10.1145/3531267}            & 2022 & $\overline{RE}$    & $AZ$              & $60$-$10$-$30$ & $Rand$ &  NDCG, R, HR, P & -                         \\ 
\hline
\multicolumn{8}{l}{$^1$ \textbf{Status} $RE$ : Replicable and Extensible; $R\overline{E}$ : Replicable but not Extensible; $\overline{RE}$ : Not Replicable nor Extensible.}\tabularnewline 
\multicolumn{8}{l}{$^2$ \textbf{Data Set} $AZ$ : Amazon \cite{amazon_dataset}; $ML$ : MovieLens 1M \cite{ml1m}; $LFM$ : LastFM \cite{lastfm-dataset}; $DB$ : DBbook2014 \cite{ktup}.}\tabularnewline
\multicolumn{8}{l}{$^3$ \textbf{Split Size} reports the percentage of data for training, validation, and test, respectively.}\tabularnewline
\multicolumn{8}{l}{$^4$ \textbf{Split Method}. $Rand$ : Random based; $Time$ : Time based.}\tabularnewline
\multicolumn{8}{l}{$^5$ \textbf{Metrics} $R$ : Recall; $HR$ : Hit Ratio $P$ : Precision; $PPC$ : Path Pattern Concentration}\tabularnewline
\end{tabular}
}
\vspace{-4mm}
\end{table}

\subsection{Methods Replication}
For each relevant paper, we analyzed the rationale of the proposed method and the characteristics of the experimental setting, as summarized in Table \ref{tab:papers}.

\texttt{PGPR} \cite{pgpr} (original source code: \url{https://github.com/orcax/PGPR}) 
was based on the idea of training a reinforcement learning (RL) agent for finding paths. 
During training, the agent starts from a user and learns to reach the correct products, with high rewards.
During inference, the agent directly walks to correct products for recommendation, without enumerating all the paths between users and products. 
The original experiments were done on four AZ data sets \cite{amazon_dataset} and on a KG built from product metadata and reviews. 

\texttt{EKAR} \cite{Song2019EkarAE} (original source code not available) 
modeled the task as a Markov decision process on the user-item-entity graph and used deep RL to solve it. 
The user-item-entity graph is treated as the environment, from which the agent gets a sequence of visited nodes and edges. 
Based on the encoded state, a policy network outputs the probability distribution over the action space. 
Finally, a positive reward is given if the agent successfully finds those products consumed by the target users in the training set. 
The novelty lays in using an LSTM for the policy network and a reward function that makes training stable and encourages agent exploration. 
Only this study included data sets from three diverse domains: movies (ML1M), music (LFM), and books (DB). 

\texttt{CAFE} \cite{cafe} (original source code: \url{https://github.com/orcax/CAFE}) 
follows the coarse-to-fine paradigm.
Given the KG, a user profile is created to capture user-centric patterns in the coarse stage. 
To conduct multi-hop path reasoning guided by the user profile, the reasoner is decomposed into an inventory of neural reasoning modules. 
Then, these modules are combined based on the user profile, to efficiently perform path reasoning. 
Original experiments followed the PGPR experimental setting (same data sets, data split, and evaluation metrics). 

\texttt{UCPR} \cite{usercentric} (original source code: \url{https://github.com/johnnyjana730/UCPR/}) 
introduces a multi-view structure leveraging not only local sequence reasoning information, but also a view of the user’s demand portfolio.
The user demand portfolio, built in a pre-processing phase and updated via a multi-step refocusing, makes the path selection process adaptive and effective. 
The original experimental setting covered the movie (ML1M) and e-commerce domains (AZ). 
This study was the only one assessing an explanation quality property, i.e., to what extent the KG relation type differs among the selected paths. 

\texttt{MLR} \cite{10.1145/3485447.3512083} (source code shared by e-mail, but external dependencies missing) 
is another RL framework that leverages both ontology-view and instance-view KGs to model multi-level user interests.
Through the Microsoft Concept Graph (MCG) \cite{10.1162/dint_a_00013}, 
the method creates various conceptual levels (e.g., Prada is an Italian luxury fashion brand). 
The reasoning is then performed by navigating through these multiple levels with an RL agent.
The authors provided the source code, but the MCG was no longer online and the provided KG dump referred only to the originally used AZ data sets. 

\texttt{PLM-Rec} \cite{10.1145/3485447.3511937}  (original source code not available), 
given a KG, extracts training path sequences under different hop constraints.
By leveraging augmentations of language features with semantics, the method obtains a series of training data sequences. 
A transformer-based decoder is then used to train an auto-regressive path language model.
This method could limit previous methods' recall bias in terms of KG connectivity. 
Only AZ data sets were used in the experiments. 

\texttt{TAPR} \cite{10.1145/3531267} (original source code not available)
proposed another path reasoning approach based on RL, characterized by the incorporation of a temporal term in the reward function. 
This temporal term guides a temporal-informed search for the agent, to capture recent trends of user's interests. 
Original results (on AZ data sets) showed a gain in utility compared to PGPR, although the model was evaluated using a random split, which is not ideal for time-aware models.

\begin{table}[!t]
\caption{Interaction and knowledge information for the two considered data sets.}
\label{tab:data-stats}
\vspace{-3mm}
\centering
\setlength{\tabcolsep}{3pt}
\parbox{.4\textwidth}{
\centering
\footnotesize
\resizebox{1\linewidth}{!}{
\begin{tabular}{lrr}
    \hline
    \textbf{Interaction} & \textbf{ML1M} & \textbf{LFM1M} \\
    \hline
    Users & 6,040 & 4,817 \\
    Products & 2,984 & 12,492 \\
    Interactions & 932,295 & 1,091,275 \\
    Density & 0.05 & 0.01  \\
    Gender (Age) Groups & 2 (7) & 2 (7) \\
    \hline
\end{tabular}}
}
\hfill
\parbox{.45\textwidth}{
\centering
\footnotesize
\renewcommand{\thesubfigure}{ii-z}
\resizebox{1\linewidth}{!}{
\begin{tabular}{lrr}
    \hline
    \textbf{Knowledge} & \textbf{ML1M} & \textbf{LFM1M} \\
    \hline
    Entities (Types) & 13,804 (12) & 17,492 (5) \\
    Relations (Types) & 193,089 (11) & 219,084 (4) \\ 
    Sparsity & 0.0060 & 0.0035 \\ 
    Avg. Degree Overall & 28.07 & 25.05 \\ 
    Avg. Degree Products & 64.86 & 17.53 \\
    \hline 
\end{tabular}}
}
\vspace{-4mm}
\end{table}

\subsection{Evaluation Protocol}
To ensure evaluation consistency and uniformity across methods, given the heterogeneous original experimental settings, we mixed replication and reproduction
\cite{10.1007/978-3-030-99736-6_37}, but use only the term ``replicability'' for convenience throughout this paper.
Specifically, we relied on the source code provided by the original authors to run their methods, and our own data and source code to (a)
pre-process the input data sets as per their requirements and (b) compute evaluation metrics based on the recommendations and reasoning paths they returned.

\noindent{\bf Data Collection.} 
We conducted experiments on two data sets: MovieLens (ML1M) \cite{ml1m} and LastFM (LFM1B) \cite{lastfm-dataset}. 
Given our interest in the fairness perspective, we selected data sets that provide (or make it possible to collect) users and providers' demographic attributes.
We therefore discarded other data sets, such as the Amazon ones \cite{amazon_dataset}, where this was not reasonably possible. 
The selected data sets are all public and vary in domain, extensiveness, and sparsity, providing novel insights on the generalizability of the replicated path reasoning methods under a common ground, with respect to their original settings (see Table \ref{tab:papers}).
For ML1M, we used the KG generated in \cite{ktup} from DBpedia, while we generated the KG from the Freebase dump extracted by \cite{kb4rec} for LFM1B. 

\noindent{\bf Data Preprocessing.} 
Concerning the ML1M data set, both gender and age sensitive attributes for the consumers, but not for the providers, were originally provided in \cite{ml1m}.
Being directors considered as movie providers in prior work \cite{DBLP:journals/umuai/BorattoFM21}, we relied on their sensitive attribute labels collected in that study.
In LFM1B, gender and age labels were attached only to a small subset of end users. 
We therefore discarded all those users whose sensitive attributes were not available. 
Given that the original papers included only data sets far smaller than LFM1B and that our preliminary experiments uncovered a low scalability for those methods\footnote{
Solving substantial scalability issues goes beyond the scope of our replicability study.}, 
we then uniformly sampled a subset of the filtered LFM1B, ensuring that users (products) had at least 20 (10) interactions. {\color{black}We will refer to this data subset as LFM1M throughout the paper and results.}
This sampling allowed us to obtain a data set size comparable to ML1M and avoid cold-start scenarios, which are not our focus.
Since providers' sensitive attributes were not attached to the original data set (in music RS, artists are commonly considered as providers), we crawled them from Freebase and released them with our study.  

Both KGs were pre-processed as performed in \cite{10.1145/3523227.3547374} to make triplets uniformly formatted. 
{\color{black}More specifically, we only consider triplets composed of a product as the entity head and an external entity as the entity tail, to obtain a common ground data set for the analysis of both knowledge-aware and path-based methods.  Considering triplets having external entities or products as the head and tail entities would have required to craft additional meta-paths (needed for path-based methods) compared to the reproduced studies, going beyond the scope of our work. In addition, to control sparsity, we removed relations having a type represented in less than 3\% of the total number of triplets}.
Concerning the user-product interactions, we discarded products (and their interactions) which are not present in the KG. 
Pre-processed data set statistics are collected in Table \ref{tab:data-stats}. 

\noindent{\bf Data Preparation and Split.} 
For each data set, we first sorted the interactions of each user chronologically. 
{\color{black}We then performed a training-validation-test split, following a time-based hold-out strategy, with the 60\% oldest interactions in the training set, the following 20\% for validation, and the 20\% most recent ones as the test set.
The aforementioned pre-processed data sets were used to train, optimize, and test each benchmarked model.
This allowed us to carry out the evaluation procedure in a realistic setting, in which the trends that might determine interaction patterns are non-stationary and evolve over time}.

\noindent{\bf Comparative Knowledge-aware Models.} 
Path reasoning methods belong to a subclass of the knowledge-aware recommendation class.
To better contextualize our study, we therefore decided to provide comparisons (when interesting) against two knowledge-aware models based on knowledge embeddings, 
namely CKE \cite{cke} and CFKG \cite{cfkg}, and a knowledge-aware model based on propagation, namely KGAT \cite{kgat}.
These three models, unable to provide reasoning paths to users, were replicated and evaluated under the same protocol\footnote{
For conciseness, we did not include non-knowledge-aware methods (e.g., BPR), which were compared against path reasoning methods under some metrics (e.g., NDCG) in studies like \cite{Balloccu2022ReinforcementRR}. Nevertheless, this is an important aspect for future work.}.
For conciseness, we do not explain their replication in detail and refer the reader to our repository. 
 
\noindent{\bf Hyper-parameter Fine-tuning.} 
Given a data set and a model, we selected the best hyper-parameters setting via a grid search that involved those hyperparameters (and their values) found to be sensitive in the original papers.
In certain cases, given our findings from preliminary experiments, we extended the grid of values to better adhere to the characteristics of the data set at hand. 
Full details on the hyper-parameters and their values in our grid search are reported in our repository. 
Models obtained via different hyper-parameter settings were evaluated on the validation set, selecting the one achieving the highest NDCG.  

\noindent{\bf Evaluation Metrics Computation.} 
Given a model and a data set, we monitored recommendation utility, beyond utility objectives, explanation quality, and both consumer and provider fairness, on recommended lists with the well-known size of $k=10$ (e.g., \cite{10.1007/978-3-030-99736-6_37}), based on the corresponding test set. 
We describe each metric in Table~\ref{tab:metrics-summary} and refer to the repository for implementation details.

Concerning recommendation utility for consumers, we monitored the Normalized Discounted Cumulative Gain (NDCG) \cite{WangWLHL13}, 
using binary relevance scores and a base-2 logarithm decay, 
and the Mean Reciprocal Rank (MRR) \cite{Craswell2009}. 
Differently from recall and accuracy, NDCG takes into account the position of the relevant products in the recommended list. 
MMR instead considers the position of the first relevant product only, giving us a perspective different than NDCG.

In our work, we focused also on four well-known beyond utility goals \cite{10.1145/2926720}. 
We monitored the extent to which the generated recommendations cover the catalog of available products (coverage). 
High coverage may increase users' satisfaction and the sales. 
Another goal, diversity, was found to be relevant for human understanding \cite{GEDIKLI2014367} and content acceptance \cite{10.1145/290941.291025}. 
We computed it as the percentage of distinct product categories in the recommended list. 
Further, serendipity measures recommendation surprise\cite{10.1145/963770.963772}. 
Given our offline setting, we compared the recommendations with those of a baseline model, i.e., a most popular recommender \cite{10.1007/978-3-540-78197-4_5}.  
The more the recommendations differ between the benchmarked and the baseline model, the higher the serendipity. 
Finally, we estimated novelty as the inverse of product popularity (as per the received ratings), assuming that products with low popularity are more likely to be surprising \cite{Zhou2010SolvingTA}. 

\begin{table}[!t]
\caption{Evaluation metrics covered in our replicability study.}
\label{tab:metrics-summary}
\vspace{-3mm}
\resizebox{1\linewidth}{!}{                     
\begin{tabular}{lllll}
\hline                           
            \textbf{Perspective}                         & \textbf{Metric}          & \textbf{Acronym} & \textbf{Range} & \textbf{Description}                                                                                                                       \\
\hline
\multirow{2}{*}{\makecell[l]{\textbf{Consumers} \\ \texttt{Utility}}} & \makecell[l]{Normalized Discounted \\ Cumulative Gain}                    & NDCG            & {[}0, 1{]}     & The extent to which the recommended products                                                \\
 &   &             &  & are useful for the user (1 means more useful).                                                \\
 & Mean Reciprocal Rank                     & MRR             & {[}0, 1{]}       & The extent to which the first recommended product                            \\
  &   &             &  &  is useful for the user (1 means more useful).                                                      \\
\hline
\multirow{4}{*}{\makecell[l]{\textbf{Consumers} \\ \texttt{Beyond} \\ \texttt{Utility}}}                                       & Coverage                      & COV             & (0, 1)         & The percentage of products overall                              \\
 &   &             &  & recommended at least once (1 means high  coverage).                                              \\
                                     & Diversity                     & DIV             & (0, 1{]}       & The percentage of product categories covered                              \\
                                      &   &             &  &  in the recommended list (1 means high diversity).\\
                                     & Novelty                       & NOV             & (0, 1)         & Inverse of the popularity of products recommended                             \\
                                              &   &             &  & to a user (1 means low popularity, so high novelty).  \\
                                     & Serendipity                   & SER             & {[}0, 1{]}     & The percentage of the recommended products not                      \\
                                              &   &             &  &  suggested also by a baseline (1 means more unexpected). \\
\hline
\multirow{8}{*}{\makecell[l]{\textbf{Consumers} \\ \texttt{Explanation} \\ \texttt{Quality}}} & Fidelity                      & FID             & {[}0, 1{]}     & The percentage of the recommended products that can                                  \\
 &   &             &  &  be explained (1 means all products can be explained).     \\
                                     & Linking Interaction Recency   & LIR             & {[}0, 1{]}     & The recency of the past interaction in the paths \\
                                                 &   &             &  & accompanying recommended products (1 means recent).     \\
                                     & Linking Interaction Diversity & LID             & (0, 1{]} & The number of distinct past interactions in the paths                                                                                                                                          \\
                                     &   &             &  & accompanying recommended products (1 means different).   \\
                                     & Shared Entity Popularity      & SEP             &   {[}0, 1{]}             &   The popularity of the shared entity in the paths                                                                                                                                        \\
                                      &   &             &  & accompanying recommended products (1 means popular).      \\
                                     & Shared Entity Diversity       & SED             &     (0, 1{]}           &  The number of distinct shared entities in the paths                 \\
                                     &   &             &  & accompanying recommended products (1 means different).                 \\  
                                     & Path Type Diversity           & PTD             &     (0, 1{]}           &  The percentage of distinct path types within paths            \\  &   &             &  &   accompanying recommended products (1 means different).     \\                                                                                                                     
                                     & Path Type Concentration       & PTC             &    {[}0, 1{]}            & The extent to which the distinct path types   \\
                                     &   &             &  & representation is equally balanced (1 means balanced).     \\  
\hline
\makecell[l]{\textbf{Providers} \\ \texttt{Utility}}                         & Exposure                      & EXP             &     {[}0, 1{]}           &    \makecell[l]{Exposure of the items of a given provider in the \\ recommended list (1 means high exposure).}\\  
\hline                                                                                                                              
\end{tabular}}
\vspace{-6mm}
\end{table}

With regard to explanation quality, we considered the proportion of explainable products in a recommended list (fidelity) \cite{10.1145/3219819.3220072}. 
In addition, we monitored reasoning paths properties concerning recency, popularity, and diversity \cite{10.1145/3477495.3532041}. 
Recent linking interactions can help the user to better catch the explanation. 
The linked interaction recency measures the recency of the past interaction presented in the explanation path,
whereas the linked interaction diversity monitors how many distinct past interactions are present.
The second perspective is related to shared entities, assuming that more popular shared entities have a higher chance of being familiar to the user. 
The shared entity popularity measures the popularity (node degree in the KG) of the shared entity in an explanation path. 
Conversely, the shared entity diversity monitors the distinct shared entities. 
Finally, path type diversity focuses on how many distinct path types are included. 
Path type concentration monitors whether path types are equally balanced. 

With the increasing importance received by fairness, we also assessed fairness with respect to a notion of demographic parity \cite{10.1007/978-3-030-99736-6_37,DBLP:conf/sigir/GomezZBSM21}. 
For consumer fairness, given a metric, we computed the average value of that metric for each demographic group and monitored the absolute pairwise difference between groups\footnote{
Our data includes sensitive attributes pertaining to the gender (Male, Female) and age (Under 18, 18-24, 25-34, 35-44, 45-49, 50-55, 56$+$), as per the data set labels.
}.
Concerning provider fairness, we computed the average exposure given to products of providers in a given demographic group \cite{DBLP:journals/umuai/BorattoFM21}.
Again, we finally computed the average absolute pairwise difference between provider groups. 

\section{Experimental Results}
Our study aimed to investigate multiple evaluation perspectives of path reasoning methods, by answering to the following research questions: 

\vspace{-1.5mm}
\begin{enumerate}[label=\textbf{RQ\arabic*},leftmargin=10mm]
    \item Do path reasoning methods trade recommendation utility and/or beyond utility objectives for explanation power? 
    \item To what extent can path reasoning methods produce explanations for all the recommended products, depending on the recommended list size? 
    \item How does the quality of the selected paths vary among path reasoning methods, based on the path type and characteristics?
\vspace{-4mm}
\end{enumerate}

\subsection{Trading Recommendation Goals for Explanation Power (RQ1)}
\label{sec:rq1}
In a first analysis, we investigated whether there exists any substantial difference in recommendation utility and beyond utility objectives  between the considered path reasoning methods (PGPR, CAFE, UCPR) and relevant knowledge-aware but not explainable methods (KGAT, CKE, CFKG). {\color{black} 
To assess statistical significance, t-tests were carried out for each metric, considering the two categories of methods as the two separate groups, under each data set. This allowed us to discern behavior also with respect to the sparsity of the KG.
Although the total sample size (six methods) is rather small, this setup made it possible to notice some preliminary characteristics of the two method classes. Still, further studies on a broader set of methods should be run to assess results generalizability more. 
P-values obtained for each test are reported in Table~\ref{tab:rq1-pvalue}.}

\begin{figure*}[!t]
\centering
\includegraphics[width=.75\textwidth]{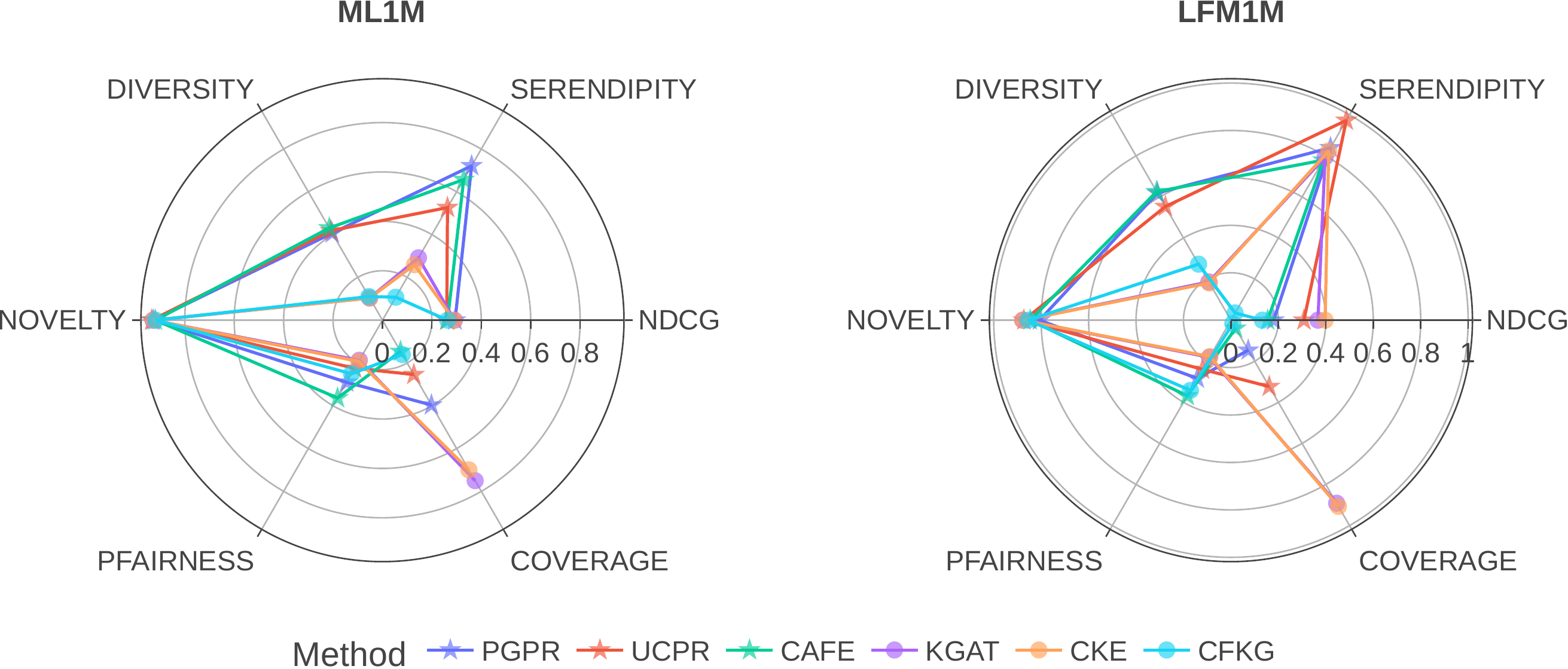}
\vspace{-3mm}
\caption{Comparison on recommendation utility and beyond utility goals [RQ1].} 
\label{fig:rq1-radar}
\vspace{2mm}
\end{figure*}

Figure~\ref{fig:rq1-radar} depicts our evaluation results in terms of utility (NDCG, MMR) and beyond utility goals (serendipity, diversity, novelty, and coverage). We also report provider fairness estimates, whereas we will discuss consumer fairness later on. Even though Figure~\ref{fig:rq1-radar} makes the comparison easier, we refer to Table \ref{tab:rq1} for specific values. 
Concerning recommendation utility, path reasoning methods achieved comparable scores (0.26 to 0.28 NDCG; 0.18 to 0.21 MRR) to knowledge-aware non-explainable methods (0.26 to 0.29 NDCG; 0.21 to 0.23 MRR) in ML1M. 
These observations did not hold in LFM1M, where path reasoning methods led to recommendations of lower utility (0.15 to 0.34 NDCG; 0.09 to 0.27 MMR) than knowledge-aware baselines (0.13 to 0.40 NDCG; 0.10 to 0.34 MMR).
{\color{black} However, in both cases and under both data sets, no statistical differences were found in terms of recommendation utility between method classes (all p-values were greater than $0.05$).
}

\begin{table}[!t]
\vspace{-2mm}
\caption{Metric scores for recommendation utility and beyond utility goals [RQ1].}
\vspace{-3mm}
\resizebox{1\linewidth}{!}{
\begin{tabular}{l|l|l|l|l|l|l|l||l|l|l|l|l|l|l}
\hline                       
\multicolumn{1}{c}{\textbf{Method}} & \multicolumn{7}{c}{\textbf{ML1M}}                                              & \multicolumn{7}{c}{\textbf{LFM1M}}                                             \\
                          & NDCG $\uparrow$ & MMR $\uparrow$ & SER $\uparrow$ & DIV $\uparrow$ & NOV $\uparrow$ & PF$^1$ $\downarrow_0$ & COV $\uparrow$ & NDCG $\uparrow$ & MMR $\uparrow$ & SER $\uparrow$ & DIV $\uparrow$ & NOV $\uparrow$ & PF$^1$ $\downarrow_0$ & COV $\uparrow$ \\
\hline
\texttt{CKE}  & \textbf{0.29}    & \textbf{0.23}    & 0.26 & 0.10      & \textbf{0.93}    & 0.19          & \underline{0.70} &
                \textbf{0.40}    & \textbf{0.34}    & \underline{0.82} & 0.18 & 0.88      & 0.18          & \textbf{0.91}    \\
\texttt{CFKG} & \underline{0.26} & 0.21             & 0.11 & 0.11      & 0.92 & 0.25      & 0.16          & 0.13             &
                            0.10 & 0.04             & 0.27 & 0.86      & \underline{0.34} & 0.02  \\
\texttt{KGAT} & \textbf{0.29}    & \textbf{0.23}    & 0.29 & 0.10      & \textbf{0.93}    & 0.19          & \textbf{0.75}    &                        \underline{0.37} & \underline{0.31} & 0.79 & 0.19      & \textbf{0.88} & 0.18 & \underline{0.89} \\
\hline
\texttt{PGPR} & 0.28             & 0.21             & 0.78 & 0.42      & 0.93          & 0.27 & 0.42 & 0.31 & 0.25 & 0.81 & 0.54 & 0.82 & 0.32 & 0.20         \\
\texttt{UCPR}                      & \underline{0.26} & 0.20 & 0.53 & \underline{0.42} & \textbf{0.93} & 0.22 & 0.25         & 0.34 & 0.27 & \textbf{0.94} & \underline{0.57} & \underline{0.87} & 0.22 & 0.41 \\
\texttt{CAFE}                      & \underline{0.26} & 0.18 & \underline{0.63} & \textbf{0.44} & \textbf{0.93} & \textbf{0.36} & 0.21         & 0.15 & 0.09 & 0.75 & \textbf{0.58} & 0.84 & \textbf{0.36} & 0.11 \\ 
\hline
\multicolumn{15}{l}{For each dataset: best result in \textbf{bold}, second-best result \underline{underlined}. $\; \;$ $^1$ \textbf{Metrics} $PF$: Provider Fairness.}\tabularnewline
\end{tabular}
}
\label{tab:rq1}
\end{table}

\begin{table}[!t]
\vspace{-2mm}
\caption{T-test p-values to assess statistically significant differences between path-based and knowledge-aware methods across evaluation metrics [RQ1].}
\vspace{-3mm}
\centering
\resizebox{.6\linewidth}{!}{
\begin{tabular}{l|r|r|r|r|r|r|r}
\hline
\textbf{Data Set$^1$} & NDCG & MMR & SER & DIV & NOV & PF & COV \\
\hline
\textbf{ML1M} & 0.33 & 0.08            & \textbf{0.01} & \textbf{0}     & 0.422  & 0.21      & 0.33    \\
\hline
\textbf{LFM1M} & 0.77             & 0.65             & 0.38 & \textbf{0}      & 0.16          & 0.38 & 0.34         \\ 
\hline
\multicolumn{8}{l}{P-values below $0.05$ are reported in \textbf{bold}.} \tabularnewline
\end{tabular}
}
\label{tab:rq1-pvalue}
\vspace{-3mm}
\end{table}

\begin{figure*}[!t]
\centering
\includegraphics[width=.8\textwidth]{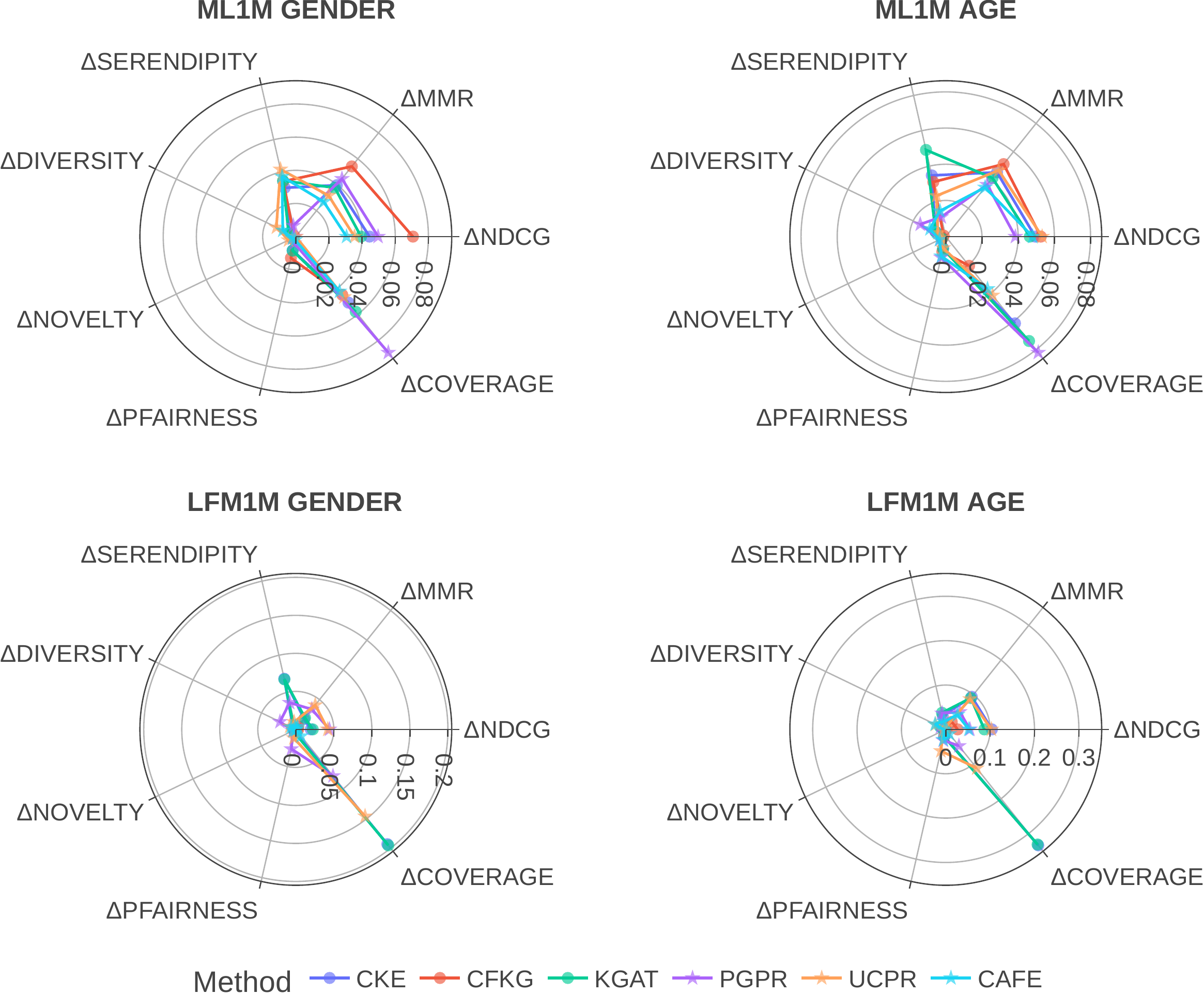}
\vspace{-3mm}
\caption{Disparate impacts between groups (gender and age) on recommendation utility, beyond utility objectives, and provider fairness. The lower it is, the fairer [RQ1].} 
\label{fig:rq1-fairness-radar}
\vspace{-4mm}
\end{figure*}

With regard to beyond utility objectives, path reasoning methods achieved substantially higher serendipity (0.53 to 0.78 ML1M; 0.75 to 0.94 LFM1M) than knowledge-aware non-explainable baselines (0.11 to 0.29 ML1M; 0.04 to 0.82 LFM1M). 
Interestingly, it was confirmed that path reasoning methods tend to perform worse in LFM1M than ML1M. 
{\color{black} Furthermore, our statistical tests show that, on ML1M, the two method classes performed differently in terms of serendipity (p-value equal to 0.011), with path reasoning methods showing higher serendipity than the knowledge aware methods, on average.
Similar patterns were found for diversity. Path reasoning methods led to a higher diversity on average (respectively 
for path-reasoning and knowledge-aware methods, 0.56 and 0.21 on LFM1M; 0.43 and 0.10 on ML1M), on both ML1M and LFM1M (p-value $\approx 0$). }
{\color{black} Conversely, on coverage, the best path reasoning method showed a decrease of 44\% on ML1M and 54.9\% on LFM1M than the best knowledge-aware method. This might be due to the low number of paths in the KG available to the path-reasoning methods. Except for CFKG, there is evidence that knowledge-aware methods should be preferred in case someone aims to optimize for coverage.
}
Finally, novelty scores were similar between the two classes of methods.

From a provider fairness perspective, path reasoning methods led to a fairer exposure of provider groups (0.22 to 0.36 ML1M; 0.22 to 0.36 LFM1M), compared to the other family (0.19 to 0.25 ML1M; 0.18 to 0.34 LFM1M). 
Surprisingly, CFKG reported the second best provider fairness score, despite of its low recommendation utility.
On the other hand, consumer fairness estimates according to the considered evaluation metrics\footnote{
Differences between the demographic groups achieving the best and worst score on avg. were all statistically significant under t-test (if applicable) or h-test otherwise.
} are collected in Figure \ref{fig:rq1-fairness-radar}. 
Being the patterns comparable across data sets and demographic groups, we describe only the results obtained in ML1M and the gender groups.
For the latter, all the models presented some yet low levels of unfairness in term of utility (both NDCG and MRR). 
Path reasoning methods (PGPR, UCPR, CAFE) achieved higher levels of unfairness on coverage, respectively 0.084, 0.045, 0.032, compared to baseline methods (KGAT reached the highest coverage unfairness, with 0.06). 
For the other metrics, we uncovered small yet comparable differences.

\vspace{-1mm}
\hlbox{Findings RQ1}{
Path reasoning methods trade recommendation utility and coverage for explanation power, especially in LFM1M.
Conversely, they resulted in higher estimates on other beyond utility objectives and provider fairness than knowledge-aware non-explainable baselines. 
}
\vspace{-1mm}

\subsection{Producing Explanations for All Recommended Products (RQ2)}
In a second analysis, we were interested in understanding the extent to which path reasoning methods can produce explanations 
for all the recommended products across recommended lists of different sizes. 
This property, fidelity, is essential for a method which yields reasoning paths (and produces explanations).  
Table \ref{tab:rq3} shows fidelity scores for the two data sets on lists of size 10, 20, 50 and 100. 

Concerning PGPR, paths were attached to almost every product of the recommended list, until the size of 50. 
Under a size of 100, fidelity remarkably decreased to 78\%. This decay in fidelity was exacerbated in LFM1M.
In the latter data set, already with a size of 20, only 74\% of recommended products were explained.
With regard to UCPR, we observed a similar but reversed pattern, compared to PGPR, on the two considered data sets. 
UCPR was challenged to produce explanations even under a size of 10 on ML1M (only 61\% of products were explained).
Conversely, in LASTFM, the same model obtained a higher fidelity than PGPR, with 99\% of explained products under a size of 10. 
Surprisingly, CAFE was able to provide a reasoning path for each recommended product until a list of size 100. 
It should however be noted that, to make this happen, the list size must be specified in advance during training. 
CAFE indeed automatically adapts the size of the neighbourhood to search around, according to the list size.
Hence, this method would be the best choice when the list size is known in advance, constant, or up to a certain limit. 
Although fidelity could be controlled, doing this led to a smaller NDCG for CAFE (see Section \ref{sec:rq1}). 

\vspace{-2.5mm}
\hlbox{Findings RQ2}{
Path reasoning methods show very different patterns in terms of fidelity.
CAFE's fidelity is high and stable across data sets and recommended list sizes. 
On the other hand, PGPR provides higher but rapidly decaying fidelity in ML1M than LFM1M, viceversa for UCPR. 
}

\begin{table}[!t]
\centering
\caption{Explanation fidelity analysis across cut-offs $k=\{10,20,50,100\}$ [RQ2].}
\label{tab:rq3}
\vspace{-2mm}
\setlength{\tabcolsep}{6pt}
\resizebox{.6\linewidth}{!}{
\begin{tabular}{l|rrrr|rrrr}
\hline
\textbf{Method} & \multicolumn{4}{c}{\textbf{ML1M}} & \multicolumn{4}{c}{\textbf{LFM1M}} \\
       & 10  & 20  & 50 & 100 & 10 & 20  & 50  & 100 \\
\hline
\texttt{PGPR}  & 1.00 & 0.99 & 0.99 & 0.78    & 0.98 & 0.74 & 0.31 & 0.15    \\
\texttt{CAFE}  & 1.00 & 1.00 & 1.00 & 1.00    & 1.00 & 1.00 & 1.00 & 1.00 \\
\texttt{UCPR}  & 0.61 & 0.34 & 0.14 & 0.07  & 0.99    & 0.98    & 0.68    & 0.35\\
\hline
\end{tabular}}\\
\vspace{-3mm}
\end{table}

\subsection{Differences on Explanation Quality (RQ3)}
\label{sec:rq2}
In a final analysis, we investigated how the quality of the selected paths (and so of the resulting explanations) varied based on the path characteristics. 
To this end, Table \ref{tab:rq2} collects seven explanation path quality perspectives (LIR, LID, SEP, SED, PTD, PTC, PPC) for each reasoning path method and data set. 

Concerning the recency dimension, we did not observe any substantial difference in linked interaction recency among the three methods, 
with the maximum (minimum) value 0.44 (0.34) achieved by PGPR (CAFE) on ML1M (similarly on LFM1M).
Whereas, in terms of linked interaction diversity, it can be interestingly noted that PGPR (0.84 ML1M; 0.77 LFM1M) and UCPR (0.82 ML1M; 0.84 LFM1M) led to higher diversity than CAFE. 

Moving to the popularity perspective and, the shared entity popularity in particular, 
CAFE was able to obtain the highest SEP in both data sets (0.75 ML1M; 0.77 LFM1M), 
meaning that it had the tendency to yield paths with more popular shared entities. 
Compared to CAFE, PGPR and UCPR showed instead substantially lower values.
Estimates on SED were very high (0.92 to 1 ML1M; 0.78 to 0.98 LFM1M) for all the methods.
These methods had hence the ability to include a good variety of shared entities in their reasoning paths.

Patterns regarding path types were particularly interesting. 
Specifically, both path type diversity and path type concentration were higher in CAFE and UCPR than PGPR. 
This highlights that the explanations of the former were, on average, richer in terms of path types (e.g., starred by, directed by), 
while PGPR's explanations were limited, on average, to a narrow set of different path types.

\begin{table}[!t]
\centering
\vspace{-2mm}
\caption{Explanation quality analysis [RQ3].}
\label{tab:rq2}
\vspace{-2mm}
\resizebox{1\linewidth}{!}{
\begin{tabular}{l|l|l|l|l|l|l|l||l|l|l|l|l|l|l}
\hline
\textbf{Model} & \multicolumn{7}{c}{\textbf{ML1M}}                & \multicolumn{7}{c}{\textbf{LFM1M}}               \\
      & LIR $\uparrow$ & LID $\uparrow$ & SEP $\uparrow$& SED $\uparrow$& PTD $\uparrow$& PTC $\uparrow$& PPC $\uparrow$& LIR $\uparrow$& LID $\uparrow$& SEP $\uparrow$& SED $\uparrow$& PTD $\uparrow$& PTC $\uparrow$& PPC $\uparrow$\\
\hline
\texttt{PGPR}  & \textbf{0.44} & \textbf{0.84} & \underline{0.43} & \underline{0.99} & 0.12 & \underline{0.03} & 0.12  & \textbf{0.49} & \underline{0.77} & 0.61 & \underline{0.94} & \underline{0.30} & 0.05 & 0.24   \\
\texttt{CAFE}  & 0.34 & 0.16 & \textbf{0.75} & \textbf{1.00} & \textbf{0.33} & \textbf{0.73} & \textbf{0.37}   & \textbf{0.49} & 0.25 & \textbf{0.77} & \textbf{0.98} & 0.25 & \textbf{0.62} & \textbf{0.50}   \\
\texttt{UCPR}  & \underline{0.40} & \underline{0.82} & 0.35 & 0.92 & \underline{0.24} & 0.01 & \underline{0.24}   & \textbf{0.49} & \textbf{0.84} & \underline{0.67} & 0.78 & \textbf{0.42} & \underline{0.24} & \underline{0.34}\\
\hline 
\multicolumn{15}{l}{For each dataset: best result in \textbf{bold}, second-best result \underline{underlined}.}\tabularnewline
\end{tabular}
}
\end{table}

\begin{figure*}[!t]
\centering
\includegraphics[scale=0.42]{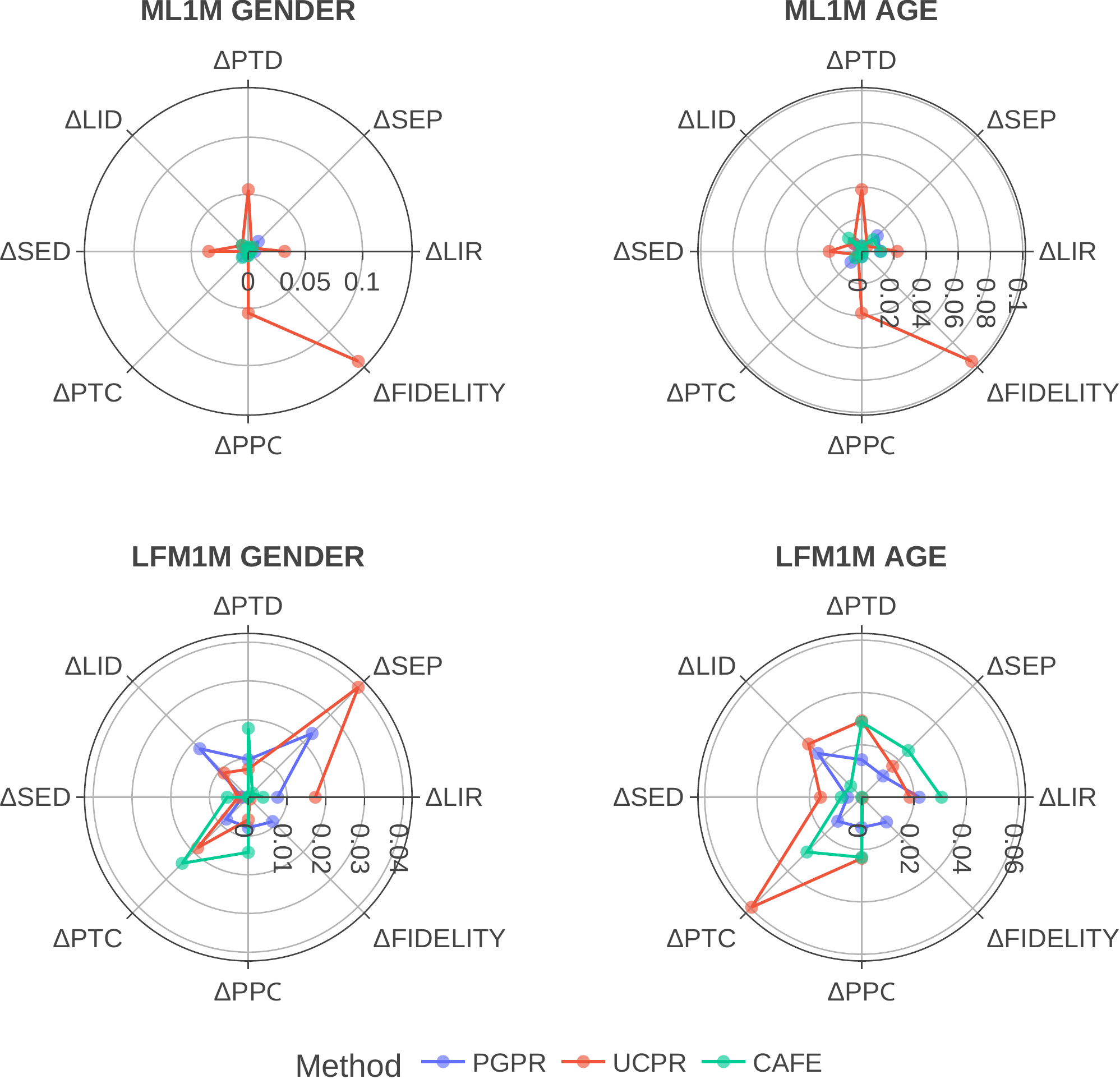}
\vspace{-2mm}
\caption{Disparate impacts between groups (gender and age) on explanation quality 
pertaining to recency, popularity, and diversity. The lower it is, the fairer [RQ3].} 
\label{fig:rq2-fairness-radar}
\vspace{-3mm}
\end{figure*}

Figure \ref{fig:rq2-fairness-radar} depicts pairwise differences in the average score of a given metric between demographic groups (consumer fairness). 
Again, under the same conditions of RQ1, all the differences were statistically valid. 
For conciseness, we discuss only the results for gender groups. 
PGPR and CAFE showed very low unfairness estimates across all metrics, 
with all scores lower than 0.01 in UCPR and 0.02 in PGPR on both data sets. 
Differently, UCPR emphasized unfairness on PTD, SED and PPC. 
Remarkably, the strongest disparate impact was reported on FID (0.12), PTD and PPC (0.05) for both data sets, and SED (0.03). {\color{black} We conjecture that these estimates of unfairness could be driven by data imbalance across the different demographic groups.}

\vspace{-1mm}
\hlbox{Findings RQ3}{
Path reasoning methods often yield substantially different paths in terms of recency, popularity, and diversity.  
Although they exist, no remarkable disparate impacts on explanation quality were found. 
}
\vspace{-3mm}

\section{Discussion and Conclusion}
In this section, we connect the main findings coming from the individual experiments 
and present the implications and limitations of our replicability study.

In the first analysis, we analyzed how methods able to produce explanations (path reasoning) compared against knowledge-aware non-explainable baselines in terms of utility, beyond utility objectives, and provider fairness. {\color{black}Results show that these two classes of methods diverge slightly in terms of utility, although explanations may have a persuasive effect which could not be captured offline. Further studies should investigate how explanations impact user decisions and consequently utility. Considering beyond accuracy objectives, we observed that path reasoning methods, due to their internal mechanics, tend to favor serendipity and diversity.} At the same time, the methodological decisions made to produce explanations make the methods more sensible to the KG structure, consequently resulting in low product coverage for the benchmarked methods. Results also show that all methods (including baselines) emphasize some levels of unfairness in almost all perspectives, especially utility. Our study calls for debiasing methods that consider multiple perspectives in the knowledge-aware setup. 

In the second analysis, we analyzed whether path reasoning methods can produce explanations across various recommended list sizes. What emerged is that some models (e.g., UCPR) are more sensible to the data and KG composition, which influence their capability of producing reasoning paths even under short recommended lists. {\color{black} This limitation could be avoided by making specific model design choices. For example, CAFE operates with a pre-defined search space for each user to deliver reasoning paths for each recommended product (although this design choice might affect recommendation utility).}

In the last analysis, we went beyond the ability of just producing reasoning paths, focusing on their quality. 
Several studies highlighted the benefits of explanations \cite{Tintarev2007}.
Recent studies also showed that path properties (e.g., recency, popularity, and diversity) can influence the user perception of explanations \cite{10.1145/3477495.3532041}. 
Results show that not all of these goals can be met at the same time. 
For instance, PGPR fails to produce diverse explanations in ML1M, whereas CAFE yields explanations based on a tiny set of past user interactions. Future studies should address this aspect through in- and post-processing methods and look at other explanation perspectives (e.g., persuasiveness, trust, and efficiency).

Overall, our analyses showed that replicating research in this area is still a challenging task.
In future work, we plan to explore in detail the impact of KG characteristics on the considered perspectives, as well as
devise novel path reasoning methods robust to the KG structure and effective on multiple objectives. 

\bibliographystyle{splncs04}
\bibliography{bibliography}

\end{document}